\documentclass[notitlepage,a4paper,aps,prd,onecolumn,superscriptaddress,nofootinbib,groupedaddress]{revtex4}

\usepackage[T1]{fontenc}
\usepackage[utf8]{inputenc}
\usepackage{amsmath,amssymb,amsfonts,amsthm,mathtools,bm}
\usepackage{graphicx,float}
\usepackage{booktabs}
\usepackage{multirow}
\usepackage{physics}
\usepackage{enumitem}
\usepackage{xcolor}
\usepackage{hyperref}
\hypersetup{colorlinks=true,linkcolor=blue,citecolor=blue,urlcolor=blue}
\allowdisplaybreaks

\newcommand{\Order}{\mathcal{O}}
\newcommand{\KY}{K_Y}

\newcommand{\Qtop}{Q_{\rm top}}
\newcommand{\rh}{r_{\rm h}}
\newcommand{\rph}{r_{\rm ph}}
\newcommand{\risco}{r_{\rm ISCO}}
\newcommand{\OISCO}{\Omega_{\rm ISCO}}
\newcommand{\Oph}{\Omega_{\rm ph}}
\newcommand{\lamph}{\lambda_{\rm ph}}

\begin{document}

\title{Geometrically Regular Black Holes with Hedgehog Scalar Hair}

\author{Sebastian Bahamonde}
\email{sbahamondebeltran@gmail.com}
\affiliation{Cosmology, Gravity, and Astroparticle Physics Group, Center for Theoretical Physics of the Universe, Institute for Basic Science (IBS), Daejeon 34126, Korea}

\begin{abstract}
We study a simple theory based on general relativity, minimally coupled to a constrained scalar triplet and to an auxiliary non-propagating three-form sector. Within a spherically symmetric hedgehog ansatz, the theory admits a continuous exact family of asymptotically flat geometrically regular black holes. For a simple choice of kinetic function, the solutions possess a de Sitter core and approach Schwarzschild with the first correction appearing only at order $r^{-4}$. We analyse their horizon structure, thermodynamics, and main strong-field properties. The black holes carry topological scalar hair and a continuous secondary parameter, but no scalar charge. The regularity established here is geometric: the curvature invariants remain finite, although the matter sector is not completely smooth at the centre.
\end{abstract}
\maketitle

\section{Introduction}
\label{sec:intro}

Regular black holes and black-hole hair are two longstanding themes in classical gravity. The first concerns whether the central curvature singularity of a black hole can be replaced by a regular core~\cite{Lan:2023ypm,Carballo-Rubio:2025bpr}. The second concerns which matter sectors can support non-trivial black-hole solutions in asymptotically flat spacetime despite the strong restrictions imposed by no-hair results~\cite{Bekenstein:1995un,Herdeiro:2015waa,Yazadjiev:2025wha}. A natural question at the intersection of these subjects is whether one can obtain an exact, asymptotically flat, geometrically regular black hole from a comparatively simple theory.

Several routes towards regular black holes have been explored. One possibility is to modify the gravitational
sector itself. Another is to retain general relativity and introduce effective matter sources, most notably of nonlinear-electrodynamic type; broad overviews are given in Refs.~\cite{Lan:2023ypm,Carballo-Rubio:2025bpr}. At a more geometric or effective level, regular Schwarzschild interiors with a Schwarzschild exterior and no primary hair have also been constructed, and their time-dependent extensions have been used to analyse the kinematics of singularity formation during collapse~\cite{Ovalle:2025pue,Ovalle:2026lxb}. These constructions show that regular cores and asymptotic flatness can coexist. At the same time, regularity of the background geometry does not by itself guarantee dynamical viability. A recent analysis showed that a broad class of nonsingular black holes supported by nonlinear electrodynamics exhibits angular Laplacian instabilities near the centre, so that geometric regularity of the background need not be preserved under linear perturbations~\cite{DeFelice:2024seu}. This provides an additional reason to examine other matter sectors. 

Scalar-tensor theories provide another useful comparison. Exact asymptotically flat black-hole branches are known in shift-symmetric Horndeski models~\cite{Kobayashi:2014eva,Babichev:2017guv}, while $SO(N)$ multi-Galileons furnish a formal framework for rotationally invariant scalar multiplets~\cite{Aoki:2021mok}. These examples show that non-trivial scalar structure can be compatible with analytically controlled black-hole solutions, but they do not answer the narrower question addressed here: can Einstein gravity admit an exact asymptotically flat regular black hole supported by a simple minimally coupled scalar sector with genuine angular structure? In such a setting, both the background mechanism and the perturbation problem are qualitatively different from the electrically or magnetically supported nonlinear-electrodynamic case.

If, in addition, one wants the matter fields themselves to carry genuine angular structure, the problem becomes more restrictive. For a single real scalar, explicit angular dependence is generically incompatible with an exactly spherically symmetric metric, even before issues of regularity or asymptotic flatness arise. This suggests turning instead to scalar multiplets, for which the angular dependence can be absorbed into an internal index. The same mismatch between angular gradients in the matter sector and exact spherical symmetry of the metric arises more generally for local scalar profiles with nonzero angular derivatives; the simple ansatz used below is only the most transparent way of displaying it.

For scalar triplets, the natural starting point is the global monopole~\cite{Barriola:1989hx}. An $O(3)$ multiplet in a hedgehog configuration carries topological charge and non-trivial angular structure, but its energy density falls only as $1/r^2$, so the spacetime develops a solid-angle deficit rather than ordinary asymptotic flatness. The corresponding geometries and their black-hole counterparts have been studied extensively~\cite{Harari:1990cz,Rhie:1991zr,Nucamendi:2000cs,Bronnikov:2002vy,Tamaki:2003vb}, including noncanonical variants~\cite{Liu:2009cb,Prasetyo:2015lzd,Carames:2023dvu}. Related mechanisms also appear in Einstein--Skyrme theory~\cite{Luckock:1986tr,Droz:1991cx,Canfora:2025roy}, in the generalised hedgehog construction~\cite{Canfora:2013osa}, and in Einstein--sigma-model and Einstein--Skyrme systems with non-spherical horizon topology~\cite{Astorino:2017wzl,Astorino:2018zvf,Henriquez-Baez:2022msz,Henriquez-Baez:2024qft}. Beyond scalar triplets, closely related hedgehog-type constructions also arise in genuinely non-Abelian sectors, where meron configurations can support regular black holes in generalised gauge theories~\cite{Diez:2026bmp}. A useful comparison is provided by a gravitating Goldstone model that admits asymptotically flat finite-mass scalar black holes and solitons without a deficit angle, albeit numerically and in a different matter theory~\cite{Radu:2011uf}. These results show that global-monopole asymptotics are not inevitable for scalar triplets, but a simple exact asymptotically flat regular-black-hole realisation has remained elusive.

We consider Einstein gravity coupled to a constrained $SO(3)$ scalar triplet and to a non-propagating three-form sector. The three-form introduces no new local propagating degrees of freedom and serves only to promote the overall density scale of the triplet kinetic sector to an integration constant, thereby allowing a continuous exact black-hole family within one fixed theory. On the hedgehog branch, the field equations close exactly: the modulus is fixed, spherical symmetry enforces the one-function form of the metric, and the Einstein equations reduce to a first-order equation for the mass function. Particular kinetic functions then generate an exact family of asymptotically flat geometrically regular black holes. Within this family, the cubic profile
\(K(Y)\propto [Y/(Y+\mu_\star^2)]^3\)
is the first simple choice for which the asymptotic departure from Schwarzschild is delayed until order \(r^{-4}\), rather than appearing already at order \(r^{-2}\). It yields black holes with topological hair and a continuous family parameter, but no independent scalar charge. The corresponding regularity statement is therefore a geometric one: the spacetime remains regular at the level of curvature invariants, even though the fixed-modulus order parameter is not smooth at the origin.

The paper is organised as follows. In Sec.~\ref{sec:single} we discuss the obstruction encountered by a single scalar with explicit angular dependence when one seeks an exactly spherically symmetric black hole. Sec.~\ref{sec:model} introduces the constrained triplet theory and derives the field equations in spherical symmetry. In Sec.~\ref{sec:family} we construct the exact regular family and then focus on its simplest asymptotically flat member. Sec.~\ref{sec:n3} studies the physical properties of that solution, including its horizon structure, thermodynamics, and strong-field behaviour. We close in Sec.~\ref{sec:conclusions} with our main conclusions.

\section{Obstruction for a single angular scalar}
\label{sec:single}

Before introducing the triplet, it is useful to explain why it is difficult for a single real scalar with explicit angular dependence to support an exactly spherically symmetric black hole. Consider the most general static,
spherically symmetric metric in spherical coordinates:
\begin{equation}
ds^2=-A(r)\,dt^2+\frac{dr^2}{B(r)}+r^2\left(d\theta^2+\sin^2\theta\,d\varphi^2\right),\label{eq:metricAB}
\end{equation}
and the simplest angular ansatz
\begin{equation}
\phi=q\,\theta+\Psi(r).
\end{equation}
This is the angular counterpart of the linearly time-dependent profile $\phi=qt+\psi(r)$ familiar in shift-symmetric scalar-tensor theories~\cite{Babichev:2013cya,Kobayashi:2014eva}. Unlike the time-dependent case, however, $q\,\theta$ is not globally well defined as a smooth scalar on $S^2$, so the expression above should be regarded only as a local ansatz, introduced to diagnose as transparently as possible whether explicit angular dependence can be compatible with an exactly spherically symmetric metric. For this simple ansatz, the kinetic invariant depends only on $r$:
\begin{equation}
X=-\frac12\,\nabla_\mu\phi\,\nabla^\mu\phi
=-\frac12\left(B(r)\Psi'(r)^2+\frac{q^2}{r^2}\right).
\end{equation}
In a generic $k$-essence theory defined by a function $P(X)$, the energy-momentum tensor is
\begin{equation}
T_{\mu\nu}=P_X\,\nabla_\mu\phi\,\nabla_\nu\phi+P\,g_{\mu\nu}\,,
\end{equation}
where $P_X=dP(X)/dX$ and one immediately finds
\begin{equation}
T_{r\theta}=P_X\,q\,\Psi'(r).
\end{equation}
Moreover, the angular diagonal components satisfy
\begin{equation}
T^\theta{}_\theta-T^\varphi{}_\varphi=\frac{P_X\,q^2}{r^2}.
\end{equation}
For a static spherically symmetric metric one has that the Einstein tensor identically $G_{r\theta}=0$ and
$G^\theta{}_\theta=G^\varphi{}_\varphi$. Hence a single angular scalar cannot source an
exactly spherically symmetric geometry in a generic $k$-essence theory. If $\Psi'(r)\neq0$, the
obstruction appears already in the off-diagonal component $T_{r\theta}$. If $\Psi'(r)=0$,
the off-diagonal term disappears but the angular stresses remain unequal. The only exceptions are $q=0$, which eliminates the angular dependence altogether, and stealth-like solutions with $P_X=0$, in which the scalar field does not give rise to genuine hair.

In Horndeski or DHOST theories, the obstruction is even stronger because second derivatives of
the scalar produce explicit angular dependence. For the same ansatz one finds
\begin{equation}
\Box\phi
=
B\,\Psi''+
\left(
\frac{B'}{2}
+\frac{B A'}{2A}
+\frac{2B}{r}
\right)\Psi'
+\frac{q\cot\theta}{r^2},
\end{equation}
and, for example,
\begin{equation}
\nabla_\varphi\nabla_\varphi\phi
=
q\sin\theta\cos\theta
+
rB\sin^2\theta\,\Psi'(r).
\end{equation}
Therefore, any interaction containing \(\Box\phi\), \((\Box\phi)^2\), or
\(\nabla_\mu\nabla_\nu\phi\,\nabla^\mu\nabla^\nu\phi\) generically introduces explicit local angular dependence. For the ansatz above to remain compatible with an exactly spherically symmetric metric, the scalar-tensor theory would therefore have to be arranged so that all such angular contributions cancel identically in the full field equations. Although one cannot exclude the existence of highly tuned or exceptional models admitting such configurations, this would be a non-generic possibility, and it is difficult to construct physically credible solutions of this type within standard scalar-tensor frameworks. The discussion in this section should thus be understood as an obstruction for this simple single-scalar ansatz, rather than as a no-go theorem excluding every possible multi-field or higher-derivative construction. In the next section, we show how the introduction of an internal triplet index avoids this obstruction while preserving exact spherical symmetry of the metric.

\section{Constrained triplet theory and the spherically symmetric reduction}
\label{sec:model}
 A possible way to hide the angular dependence is to absorb it into an internal index. This
leads naturally to an $SO(3)$ multiplet in a hedgehog configuration,
\begin{equation}
\Phi^I=H(r)\,n^I(\theta,\varphi),
\qquad
n^I=(\sin\theta\cos\varphi,\sin\theta\sin\varphi,\cos\theta).\label{eq:hedgehog}
\end{equation}
Only after this step can the invariant become purely radial. Although $\Phi^I$ depends explicitly on $(\theta,\phi)$, the configuration is spherically symmetric in the generalised hedgehog sense: a spatial rotation can be compensated by an internal $SO(3)$ rotation. This is the same geometric mechanism that underlies hedgehog configurations in global-monopole, sigma-model and Einstein--Skyrme systems~\cite{Barriola:1989hx,Luckock:1986tr,Canfora:2013osa}.

Motivated by the obstruction above, we consider Einstein gravity coupled to a constrained scalar triplet and to a minimal auxiliary three-form sector. In four dimensions the three-form carries no local propagating degrees of freedom. This is the standard mechanism by which a top-form sector generates an integration constant in four dimensions~\cite{Aurilia1980,HenneauxTeitelboim1984}. Its purpose is only to generate the overall amplitude of the nonlinear triplet sector as an integration constant of the field equations, rather than inserting that amplitude as a fixed numerical coefficient in the action. The model is
\begin{equation}
S=\int \dd^4x\,\sqrt{-g}\left[
\frac{R}{16\pi G}
-\mathcal{C}\,\widehat K(Y)
+\lambda\left(\delta_{IJ}\Phi^I\Phi^J-\eta^2\right)
-\frac{1}{3!}\,\frac{\varepsilon^{\mu\nu\rho\sigma}}{\sqrt{-g}}\,\mathcal{A}_{\mu\nu\rho}\,\nabla_\sigma \mathcal{C}
\right],
\label{eq:action}
\end{equation}
where
\begin{equation}
Y=\frac12\,\delta_{IJ}\,\nabla_\mu\Phi^I\nabla^\mu\Phi^J .
\end{equation}
Here \(\varepsilon^{\mu\nu\rho\sigma}\) denotes the totally antisymmetric Levi-Civita symbol, normalised as \(\varepsilon^{0123}=+1\). The function \(\widehat K(Y)\) is an arbitrary function of the kinetic term (as a $k$-essence type), \(\mathcal{C}(x)\) is an auxiliary scalar that fixes the overall density scale on shell, and \(\mathcal{A}_{\mu\nu\rho}\) is a three-form potential. Written in this form, the action is manifestly covariant, and the three-form term is metric-independent.
The parameter \(\eta\) sets the symmetry-breaking scale, while the Lagrange-multiplier formulation is analogous to the nonlinear sigma-model limit in which the heavy radial mode has been integrated out~\cite{Gell-Mann:1960mvl,Coleman:1969sm,Callan:1969sn,Canfora:2013osa,Radu:2011uf}. 

Lagrange multipliers are also widely used in gravitational effective theories to impose covariant constraints at the level of the action. In mimetic gravity, for example, a constraint on a scalar gradient isolates an additional conformal degree of freedom which can behave as pressureless dust~\cite{Chamseddine:2013kea}. Related constrained-scalar constructions have been used to describe dark-sector fluids with vanishing sound speed~\cite{Lim:2010yk}, and Lagrange-multiplier constraints have also been employed in modified-gravity reconstructions of dark-energy cosmologies~\cite{Capozziello:2010uv}. More recently, analogous mimetic constraints have been applied to Abelian and non-Abelian gauge sectors, leading to black-hole solutions with electric or non-Abelian hair~\cite{Gorji:2025mimeticYM}. The multiplier in the present model plays a different and more limited role: it imposes the algebraic fixed-norm condition on the scalar triplet and removes the radial modulus from the low-energy dynamics. It does not introduce a mimetic conformal mode, but restricts the matter sector to the angular Goldstone directions of the hedgehog.

At this stage the three-form should be viewed only as a device for promoting the overall amplitude of the effective triplet sector to a solution parameter; its concrete role becomes explicit once the mass equation is derived below.
Variation with respect to $\mathcal{A}_{\mu\nu\rho}$ gives
\begin{equation}
\nabla_\sigma \mathcal{C}=0
\qquad\Longrightarrow\qquad
\mathcal{C}(x)=\rho_0,
\label{eq:Cconst}
\end{equation}
where $\rho_0$ is an integration constant rather than a coupling constant of the action. This distinction is important. At the classical level considered here, $\rho_0$ is therefore a continuous solution parameter. Any possible quantisation of the corresponding three-form flux would require additional ultraviolet input and is not assumed in the present effective description. As will become clear below, the overall amplitude of the nonlinear triplet sector sets the density scale of the source and therefore controls the ADM mass of the regular solutions. If that amplitude were inserted directly as a fixed coupling of the theory, then, once the remaining theory parameters and the regularity condition are specified, the mass of the black hole would be fixed by the theory itself rather than arising as a genuine integration constant of the solution. Physically, this is rather unnatural: varying the mass would then amount to moving between different theories, rather than between different black-hole solutions of one and the same theory, a feature that can arise in some exact closed-form constructions. The auxiliary three-form avoids this by promoting the overall amplitude to the on-shell constant $\rho_0$, while introducing no new local propagating degrees of freedom. 

Variation with respect to $\mathcal{C}$ yields
\begin{equation}
\widehat K(Y)
=\frac{1}{3!}\nabla_\sigma\!\left(
\frac{\varepsilon^{\mu\nu\rho\sigma}}{\sqrt{-g}}\,
\mathcal{A}_{\mu\nu\rho}
\right)\,.
\label{eq:Ceq}
\end{equation}
This rewriting also makes contact with the non-Riemannian volume-form approach to modified gravity and cosmology~\cite{Guendelman:1999qt,Guendelman:2015jii,Benisty:2019jqz,Benisty:2019dvu,Benisty:2020rul}. Indeed, the last term in Eq.~\eqref{eq:action} is independent of the Riemannian volume element after the explicit factor of \(\sqrt{-g}\) is cancelled:
\begin{equation}
-\int \dd^4x\,\sqrt{-g}\,
\frac{1}{3!}\frac{\varepsilon^{\mu\nu\rho\sigma}}{\sqrt{-g}}
\mathcal{A}_{\mu\nu\rho}\nabla_\sigma \mathcal{C}
=
-\frac{1}{3!}\int \dd^4x\,
\varepsilon^{\mu\nu\rho\sigma}\mathcal{A}_{\mu\nu\rho}\partial_\sigma \mathcal{C} .
\end{equation}
Up to a boundary term this may be written as
\begin{equation}
\int \dd^4x\, \mathcal{C}\,\Phi_\mathcal{A},
\qquad
\Phi_\mathcal{A}\equiv
\frac{1}{3!}\partial_\sigma
\left(
\varepsilon^{\mu\nu\rho\sigma}\mathcal{A}_{\mu\nu\rho}
\right),
\end{equation}
where \(\Phi_\mathcal{A}\) is a metric-independent scalar density constructed from the three-form potential. Thus the auxiliary sector used here is a simple particular realisation of the same general idea: a covariant integration density defined by a higher-rank antisymmetric tensor gauge field. In the present work we do not use the broader non-Riemannian volume-form framework; the three-form is introduced only to generate the integration constant \(\rho_0\) that sets the overall amplitude of the effective triplet source.

It is therefore convenient to define the effective on-shell kinetic function
\begin{equation}
K(Y)\equiv \rho_0\,\widehat K(Y).
\label{eq:Keff}
\end{equation}
From this point on, $K(Y)$ denotes this effective quantity. The fixed-norm constraint is still enforced by
\begin{equation}
\Phi^I\Phi^I=\eta^2.
\label{eq:constraint}
\end{equation}

We now allow the most general static spherically symmetric metric defined by~\eqref{eq:metricAB} and use the hedgehog ansatz~\eqref{eq:hedgehog}. The kinetic invariant is then
\begin{equation}
Y=\frac12\left(BH'^2+\frac{2H^2}{r^2}\right).
\label{eq:Yhedge}
\end{equation}
Variation with respect to the multiplier gives directly
\begin{equation}
H(r)^2=\eta^2.
\label{eq:Hconstraint}
\end{equation}
On each connected component one may therefore choose either $H=+\eta$ or $H=-\eta$; the two choices differ only by an overall sign of the triplet. In what follows we take
\begin{equation}
H(r)=\eta,
\qquad
\Phi^I=\eta\,n^I(\theta,\varphi),
\qquad
Y=\frac{\eta^2}{r^2}.
\label{eq:fixedbranch}
\end{equation}
The scalar configuration is thus fixed already by the constraint, before any choice is made for the metric.

With the constraint imposed, the energy-momentum tensor is
\begin{equation}
T_{\mu\nu}=\KY\,\delta_{IJ}\nabla_\mu\Phi^I\nabla_\nu\Phi^J-Kg_{\mu\nu},
\label{eq:Tmunu}
\end{equation}
where \(K_Y=dK(Y)/dY\), and the multiplier term does not contribute because \(\Phi^I\Phi^I-\eta^2=0\). Its mixed components are
\begin{equation}
T^t{}_t=-K,
\qquad
T^r{}_r=-K,
\qquad
T^\theta{}_{\theta}=T^\varphi{}_{\varphi}=-K+Y K_Y\,,
\label{eq:Texactgeneral}
\end{equation}
and then, in terms of the effective density and principal pressures, this implies
\begin{equation}
\rho = K,
\qquad
p_r = -\rho,
\qquad
p_t = -\rho + Y K_Y,
\label{eq:rho_pr_pt_general}
\end{equation}
so that the source is generically anisotropic already at the level of the constrained triplet theory.

For the metric~\eqref{eq:metricAB}, the corresponding Einstein tensor is
\begin{equation}
G^t{}_t=\frac{rB'+B-1}{r^2},
\qquad
G^r{}_r=\frac{B}{r}\frac{A'}{A}+\frac{B-1}{r^2}.
\label{eq:GttGrrAB}
\end{equation}
Since $T^t{}_t=T^r{}_r$ for this configuration, Einstein's equations immediately imply
\begin{equation}
\frac{A'}{A}=\frac{B'}{B}\,,
\label{eq:AprimeBprime}
\end{equation}
so that \(A/B=\text{const}\). Rescaling \(t\) sets \(A(r)=B(r)\).

Variation with respect to the triplet gives
\begin{equation}
\nabla_\mu\left(\KY\nabla^\mu\Phi^I\right)+2\lambda\Phi^I=0.
\label{eq:scalarfull}
\end{equation}
For the configuration $\Phi^I=\eta n^I$, one has $\nabla_r\Phi^I=0$, $\KY=\KY(r)$ and $\nabla^2 n^I=-2n^I/r^2$, so the multiplier is determined algebraically,
\begin{equation}
\lambda(r)=\frac{\KY}{r^2}.
\label{eq:lambdaexact}
\end{equation}
Finally, using $A=B$ in the $tt$ Einstein equation gives
\begin{equation}
rA'(r)+A(r)-1=-8\pi G r^2 K\!\left(\frac{\eta^2}{r^2}\right)\,,
\label{eq:masterA}
\end{equation}
that can be easily solved, yielding
\begin{equation}
A(r)=1-\frac{2Gm(r)}{r},
\label{eq:masterm}
\end{equation}
where we have defined a mass function determined by $K(Y)$ as:
\begin{equation}
m'(r)=4\pi r^2 K\!\left(\frac{\eta^2}{r^2}\right).
\label{eq:massdef}
\end{equation}
The field equations have thus been reduced to the single mass equation~\eqref{eq:massdef}. This relies crucially on the fixed-norm constraint: the Lagrange multiplier freezes the radial modulus, so that $H(r)$ is no longer an independent function and the scalar invariant reduces to $Y=\eta^2/r^2$. The Einstein equations then close in terms of Eq.~\eqref{eq:massdef}, and once $K(Y)$ is specified, the metric follows directly.

Integrating Eq.~\eqref{eq:massdef} still leaves an additive constant in \(m(r)\). If this constant is nonzero, it generates the usual Schwarzschild \(1/r\) term and hence a central singularity. The regular solutions of interest are therefore obtained by choosing \(m(0)=0\). At this point the role of the auxiliary three-form also becomes transparent: it promotes the overall amplitude of \(K(Y)\) to the integration constant \(\rho_0\), rather than fixing it as a coupling in the action. Regularity therefore does not collapse the model to a single ADM mass. Instead, one retains a continuous exact family of regular solutions within one fixed theory.

To illustrate the type of geometries generated by Eq.~\eqref{eq:massdef}, it is instructive to consider first a simple power-law class, for which the equation can be integrated in closed form. For \(K(Y)=\beta Y^p\) with \(\beta>0\), the exact branch \(Y=\eta^2/r^2\) gives
\begin{equation}
A(r)=1-\frac{2GM}{r}-\frac{8\pi G\beta \eta^{2p}}{3-2p}\,r^{2-2p},
\qquad
p\neq \frac32 ,
\end{equation}
whereas the marginal case \(p=\frac32\) yields the logarithmic form
\begin{equation}
A(r)=1-\frac{2GM}{r}-\frac{8\pi G\beta \eta^3}{r}\ln\!\left(\frac{r}{r_0}\right).
\end{equation}
The choice \(p=1\) reproduces the familiar global-monopole asymptotics with a solid-angle deficit. By contrast, \(p=2\) gives the asymptotically flat but singular solution
\begin{equation}
A(r)=1-\frac{2GM}{r}+\frac{8\pi G\beta\eta^4}{r^2}.
\end{equation}
Its matter sector satisfies \(\rho=\beta\eta^4/r^4\), \(p_r=-\rho\), and \(p_t=\rho\), so the triplet reproduces the same effective equation of state as Maxwell electrovac and the geometry becomes Reissner--Nordstr\"om-like, with the $1/r^2$ term set by the effective amplitude of the triplet sector rather than by a scalar Gauss-law charge. 

With these simple singular branches in mind, we now turn to a particular choice of \(K(Y)\) that preserves exact solvability and leads to a simple family whose geometric regularity can then be checked explicitly.

\section{Exact regular family and a particularly simple case}
\label{sec:family}

A convenient on-shell family of effective kinetic functions is
\begin{equation}
K_n(Y)=\rho_0\,\widehat K_n(Y),\qquad
\widehat K_n(Y)\equiv\left(\frac{Y}{Y+\mu_\star^2}\right)^n,\qquad
n>\frac32,\label{eq:Kn}
\end{equation}
where $\widehat K_n(Y)$ is the dimensionless shape function, while the overall amplitude $\rho_0$ is the integration constant generated by the auxiliary three-form sector. It is then useful to define
\begin{equation}
L\equiv \frac{\eta}{\mu_\star}
\label{eq:Ldef}
\end{equation}
that sets the core scale. Using Eq.~\eqref{eq:fixedbranch}, we find
\begin{equation}
K_n\!\left(\frac{\eta^2}{r^2}\right)=\frac{\rho_0}{\left(1+r^2/L^2\right)^n}\,,
\label{eq:Knonbranch}
\end{equation}
and then, Eq.~\eqref{eq:massdef} can be integrated yielding
\begin{equation}
m_n(r)=4\pi\rho_0L^3\int_0^{r/L}\frac{x^2\dd x}{(1+x^2)^n}
=\frac{4\pi\rho_0r^3}{3}\,{}_2F_1\!\left(\frac32,n;\frac52;-\frac{r^2}{L^2}\right),
\label{eq:mngeneral}
\end{equation}
where ${}_2F_1(a,b;c;z)$ denotes the Gaussian hypergeometric function. Then, the metric function becomes
\begin{equation}
A(r)=1-\frac{8\pi G\rho_0 r^2}{3}\,
{}_2F_1\!\left(\frac32,n;\frac52;-\frac{r^2}{L^2}\right).
\label{eq:Angeneral}
\end{equation}
It is easy to verify that near the centre:
\begin{align}
K_n&=\rho_0+\Order(r^2),
\qquad
m_n(r)=\frac{4\pi\rho_0}{3}r^3+\Order(r^5),\\
R(0)&=32\pi G\rho_0\,,\quad R_{\mu\nu}R^{\mu\nu}(0)=256\pi^2G^2\rho_0^2\,,\quad 
R_{\mu\nu\rho\sigma}R^{\mu\nu\rho\sigma}(0)=\frac{512}{3}\pi^2G^2\rho_0^2\,,
\label{eq:centerinvariants}
\end{align}
so every member of the family has a de Sitter core, with curvature scale set by $\rho_0$, namely:
\begin{equation}
A(r)=1-\frac{8\pi G\rho_0}{3}r^2+\Order(r^4).
\label{eq:dscore}
\end{equation}
Although the scalar invariant diverges as \(Y=\eta^2/r^2\) at the centre, the quantity entering the matter Lagrangian remains finite. For the family~\eqref{eq:Kn},
\begin{equation}
K_n\!\left(\frac{\eta^2}{r^2}\right)
=
\frac{\rho_0}{(1+r^2/L^2)^n}\,,
\end{equation}
and hence \(K_n\to\rho_0\) as \(r\to0\). On the one-function branch the metric determinant satisfies \(\sqrt{-g}=r^2\sin\theta\), so the local contribution of the \(K\)-sector to the action behaves near the centre as
\begin{equation}
\sqrt{-g}\,
K_n\!\left(\frac{\eta^2}{r^2}\right)
=
r^2\sin\theta\,\rho_0\left[1+O(r^2)\right].
\end{equation}
It is therefore integrable at \(r=0\). The full radial integral of the \(K\)-sector is finite for the same condition that gives finite ADM mass,
\begin{equation}
4\pi\int_0^\infty \dd r\,r^2
K_n\!\left(\frac{\eta^2}{r^2}\right)
=
\pi^{3/2}\rho_0 L^3
\frac{\Gamma\!\left(n-\frac32\right)}{\Gamma(n)}
=
M_n,
\qquad n>\frac32.
\end{equation}
Thus the divergence of \(Y\) does not make the Lagrangian contribution singular. The remaining lack of smoothness is instead associated with the fixed-modulus order parameter \(n^I=x^I/r\), which is undefined at the central point.

At large radius the mass function behaves as:
\begin{equation}
m_n(r)=M_n-\frac{4\pi\rho_0L^{2n}}{(2n-3)r^{2n-3}}+\Order(r^{-2n+1}),
\label{eq:mnasy}
\end{equation}
with ADM mass
\begin{equation}
M_n=\pi^{3/2}\rho_0L^3\frac{\Gamma\!\left(n-\frac32\right)}{\Gamma(n)}.
\label{eq:Mn}
\end{equation}
Consequently,
\begin{equation}
A(r)=1-\frac{2GM_n}{r}+\frac{8\pi G\rho_0L^{2n}}{(2n-3)r^{2n-2}}+\Order(r^{-2n}).
\label{eq:Anasy}
\end{equation}
The condition $n>3/2$ is exactly the condition for finite ADM mass.

The $n=2$ solution already has finite ADM mass, but its far field contains a $1/r^2$ correction and is therefore Reissner--Nordström-like. The $n=3$ solution is qualitatively cleaner. Its mass function is
\begin{equation}
m(r)=\frac{\pi\rho_0L^3}{2}\left[\arctan\!\left(\frac{r}{L}\right)+\frac{r}{L}\frac{r^2/L^2-1}{\left(1+r^2/L^2\right)^2}\right],
\label{eq:m3}
\end{equation}
and the ADM mass is
\begin{equation}
M=\frac{\pi^2}{4}\rho_0L^3.
\label{eq:M3}
\end{equation}
Hence the exact regular family is continuous within one fixed theory: for fixed $L$, the ADM mass varies with the integration constant $\rho_0$ rather than with a coupling of the microscopic action.
From this point on, unless stated otherwise, $A(r)$ denotes the metric function of the $n=3$ solution, which is given explicitly by
\begin{equation}
A(r)=1-\frac{4GM}{\pi r}\left[\arctan\!\left(\frac{r}{L}\right)+\frac{r}{L}\frac{r^2/L^2-1}{\left(1+r^2/L^2\right)^2}\right].
\label{eq:A3exact}
\end{equation}
Its large-radius expansion reads
\begin{equation}
A(r)=1-\frac{2GM}{r}+\frac{32GML^3}{3\pi r^4}+\Order(r^{-6})\,,
\label{eq:Aasy}
\end{equation}
and then, there is no solid-angle deficit and no $1/r^2$ term.

\section{Physical properties of the regular black hole solution}
\label{sec:n3}

\subsection{Horizon structure and thermodynamics}

For the \(n=3\) solution it is convenient to introduce the dimensionless variables
\begin{equation}
x\equiv \frac{r}{L},
\qquad
\chi\equiv \frac{GM}{L},
\label{eq:xchi}
\end{equation}
in terms of which the metric function becomes
\begin{equation}
A(x)=1-\frac{4\chi}{\pi x}\left[\arctan x+\frac{x(x^2-1)}{(1+x^2)^2}\right].
\label{eq:Adimless}
\end{equation}
Using Eq.~\eqref{eq:M3}, one may equally write $\chi=(\pi^2/4)G\rho_0L^2$. In the present formulation $\rho_0$ is an integration constant of the auxiliary three-form sector, so varying $\chi$ scans a continuous exact family within one fixed theory.
The parameter \(\chi\) measures the ratio between the gravitational scale \(GM\) and the core scale \(L\). It therefore controls whether the regular core is hidden behind an event horizon or remains exposed.

\begin{figure}[htp]
\centering
\includegraphics[width=0.66\textwidth]{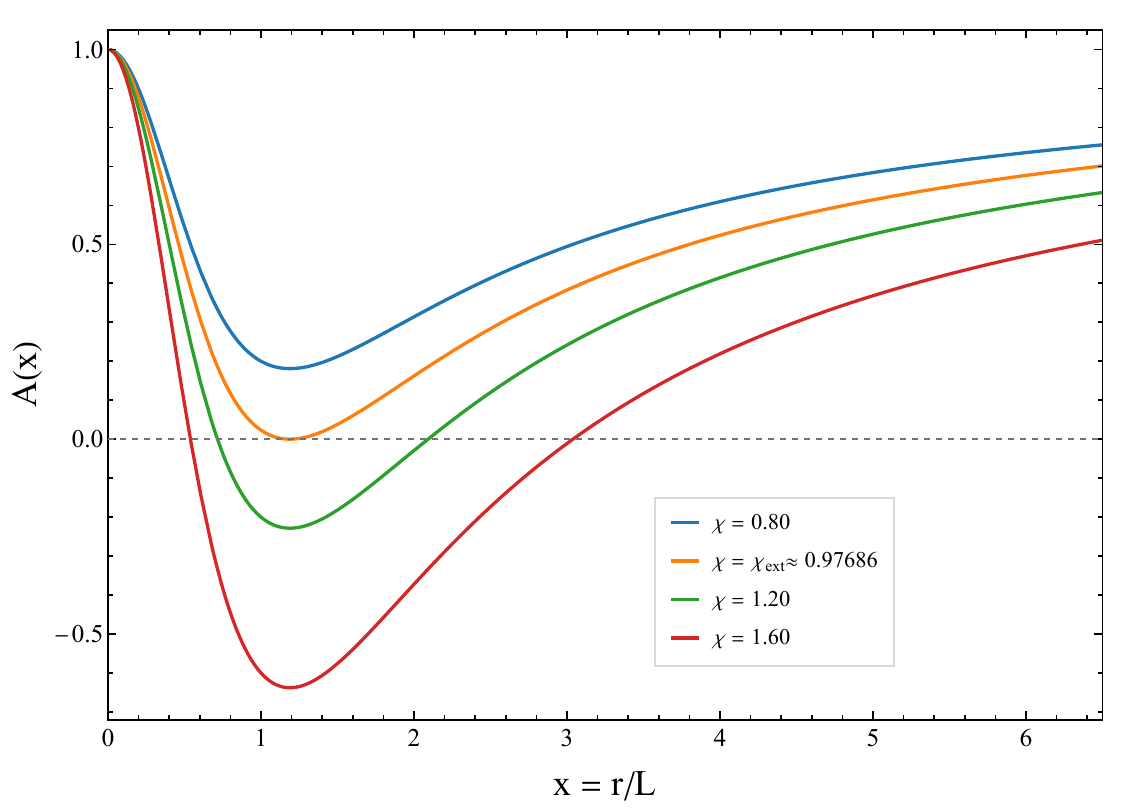}
\caption{Metric profiles \(A(x)\) of the \(n=3\) solution for representative values of \(\chi=GM/L\). The dashed line marks \(A=0\). The blue curve is horizonless, the orange curve is extremal, and the green and red curves possess two horizons.}
\label{fig:profiles}
\end{figure}

Representative profiles of \(A(x)\) are shown in Fig.~\ref{fig:profiles}. Three qualitatively distinct regimes are present. For \(\chi<\chi_{\rm ext}\), the metric function remains positive and the spacetime is horizonless. At the critical value \(\chi=\chi_{\rm ext}\), the minimum of \(A(x)\) touches zero and the solution becomes extremal. For \(\chi>\chi_{\rm ext}\), the minimum lies below zero and two horizons appear: an outer event horizon and an inner Cauchy horizon. In this sense, the parameter \(\chi\) governs the transition from a horizonless geometrically regular configuration to a genuine regular black hole. As in other regular-black-hole geometries with a non-extremal inner horizon, however, the existence of a Cauchy horizon should not be read as evidence for a dynamically stable interior. The present paper studies the exact static backgrounds only; possible mass-inflation or related inner-horizon instabilities are not analysed here~\cite{Lan:2023ypm,Carballo-Rubio:2025bpr}.

\begin{figure}[htp]
\centering
\includegraphics[width=0.66\textwidth]{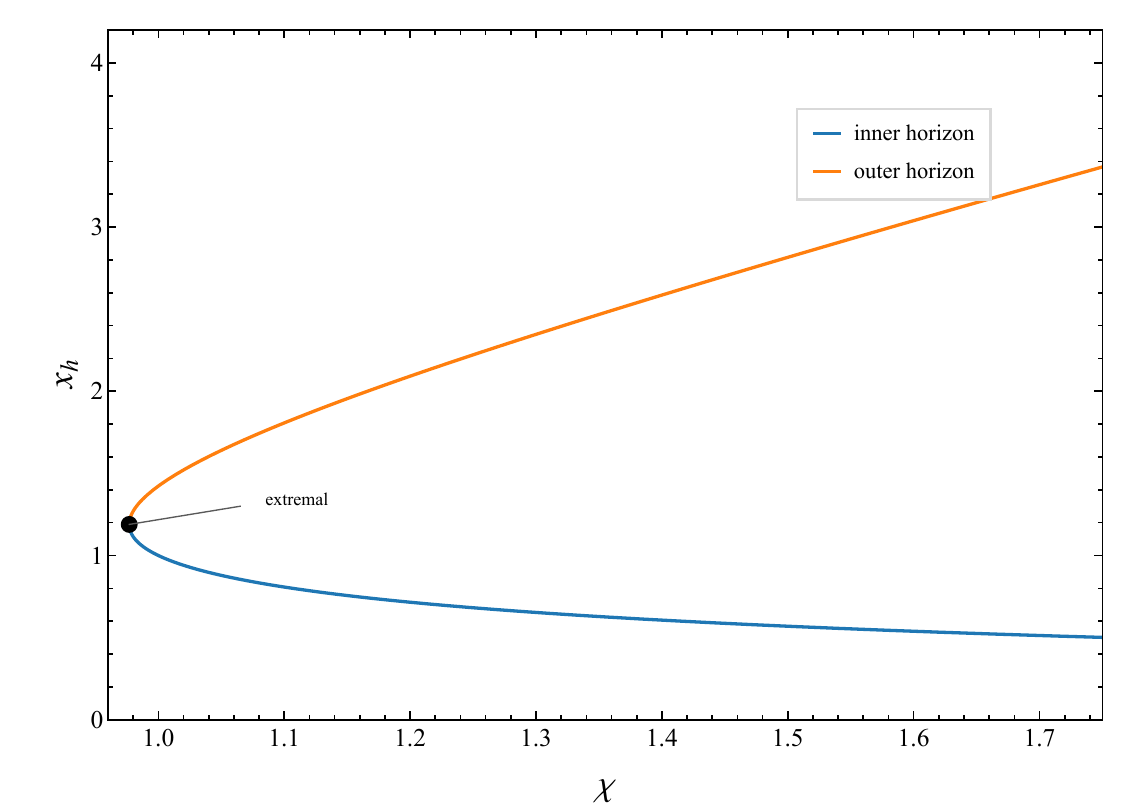}
\caption{Inner and outer horizon radii of the \(n=3\) solution in the \((\chi,x_h)\) plane, where \(\chi=GM/L\) and \(x_h=r_h/L\). The two curves meet at the extremal point \((\chi_{\rm ext},x_{\rm ext})\simeq(0.97686,1.18924)\).}
\label{fig:horizons}
\end{figure}

The extremal configuration is determined by the double-zero condition
\begin{equation}
A(x_{\rm ext})=0,\qquad A'(x_{\rm ext})=0.
\end{equation}
The corresponding horizon curves are displayed in Fig.~\ref{fig:horizons}. The outer horizon increases monotonically with \(\chi\), while the inner horizon moves inward as the black hole becomes heavier. Both merge smoothly at the extremal point, where the surface gravity vanishes.
The thermodynamic quantities follow directly from the surface gravity and the Einstein--Hilbert area law,
\begin{equation}
T_H=\frac{A'(\rh)}{4\pi}
=\frac{1-8\pi G\,\rh^2 K(\eta^2/\rh^2)}{4\pi \rh},
\qquad
S=\frac{\pi \rh^2}{G}.
\label{eq:THgeneral}
\end{equation}
The temperature therefore vanishes at extremality, as expected for a double horizon. Away from extremality, \(T_H\) is controlled by the competition between the usual Schwarzschild term \(1/(4\pi \rh)\) and the matter contribution evaluated at the horizon. The entropy retains the standard area-law form because the gravitational sector is purely Einstein--Hilbert.

For large black holes, \(GM\gg L\), the outer horizon and temperature admit the expansions
\begin{align}
\rh&\simeq 2GM-\frac{4L^3}{3\pi G^2M^2},
\label{eq:rhlargem}\\
T_H&\simeq \frac{1}{8\pi GM}\left[1-\frac{4}{3\pi}\left(\frac{L}{GM}\right)^3\right].
\label{eq:THlarge}
\end{align}
These expressions show explicitly that the regular solution approaches Schwarzschild for \(L/(GM)\ll1\), but does so from below: at fixed ADM mass the outer horizon is slightly smaller and the Hawking temperature is slightly lower. Physically, the finite core softens the geometry near the centre while leaving only suppressed corrections in the large-mass regime.

A further useful characterization of the \(n=3\) branch is obtained by using the horizon radius \(x_h=r_h/L\) itself as
the thermodynamic parameter. The horizon condition \(A(x_h)=0\) determines the mass ratio as
\begin{equation}
\chi(x_h)=\frac{\pi x_h}{4\left[\arctan x_h+\dfrac{x_h(x_h^2-1)}{(1+x_h^2)^2}\right]},
\end{equation}
while the exact Hawking temperature may be written as
\begin{equation}
L T_H(x_h)=\frac{1}{4\pi x_h}\left[1-\frac{8x_h^3}{(1+x_h^2)^3\left(\arctan x_h+\dfrac{x_h(x_h^2-1)}{(1+x_h^2)^2}\right)}\right].
\end{equation}
The heat capacity then follows parametrically from
\begin{equation}
\frac{G C_{\rm H}}{L^2}
=
\frac{d\chi/dx_h}{d(LT_H)/dx_h}\,.
\end{equation}
This parametrisation shows the thermodynamic structure of the branch.  Near extremality the temperature rises from zero as the mass increases, so the heat capacity $C_{\rm H}$ is positive. For larger black holes one recovers the familiar Schwarzschild-like behaviour with negative heat capacity. The \(n=3\) branch is therefore locally thermodynamically stable near extremality, but thermodynamically unstable in the large-mass regime. 

The discussion above is intended as a diagnostic of the exact static family. We do not attempt here a full first-law analysis including possible conjugate variables associated with the auxiliary three-form sector.


\subsection{Energy-momentum tensor, energy conditions, and topological charge}
Equation~\eqref{eq:fixedbranch} fixes the scalar invariant, so the matter sector is entirely determined by the radial dependence of the nonlinear kinetic function \(K(Y)\). Using Eqs.~\eqref{eq:Texactgeneral} and \eqref{eq:rho_pr_pt_general}, one sees that the source is generically anisotropic: the radial pressure satisfies \(p_r=-\rho\), whereas the tangential pressure differs by the additional contribution \(Y K_Y\). In particular, the radial equation of state is vacuum-like, as expected in regular de Sitter-core geometries. Regularity is therefore tied to an effective vacuum behaviour in the deep interior, while the angular stresses govern the departure from isotropy away from the centre and the transition towards the asymptotically flat region.

For the \(n=3\) solution, one obtains
\begin{align}
\rho(r) &= \frac{\rho_0}{\left(1+r^2/L^2\right)^3},
\label{eq:rho3}\\
p_r(r) &= -\rho(r),
\\
p_t(r) &= \rho_0 \frac{2r^2/L^2-1}{\left(1+r^2/L^2\right)^4}.
\label{eq:pt3}
\end{align}
These profiles show the structure of the source. At the centre,
\begin{equation}
\rho(0)=\rho_0,
\qquad
p_r(0)=p_t(0)=-\rho_0,
\end{equation}
so the matter distribution approaches an isotropic de Sitter core. This is the mechanism by which the Schwarzschild singularity is replaced by a regular interior. Away from the centre, isotropy is broken: the radial pressure remains negative, whereas the tangential pressure changes sign at
\begin{equation}
r=\frac{L}{\sqrt{2}},
\end{equation}
and becomes positive for larger radii. In the asymptotic region, the density decays as \(r^{-6}\), while the tangential pressure approaches \(2\rho\), so the source smoothly interpolates between a vacuum-like central core and a rapidly decaying anisotropic halo.

\begin{figure}[H]
\centering
\includegraphics[width=0.66\textwidth]{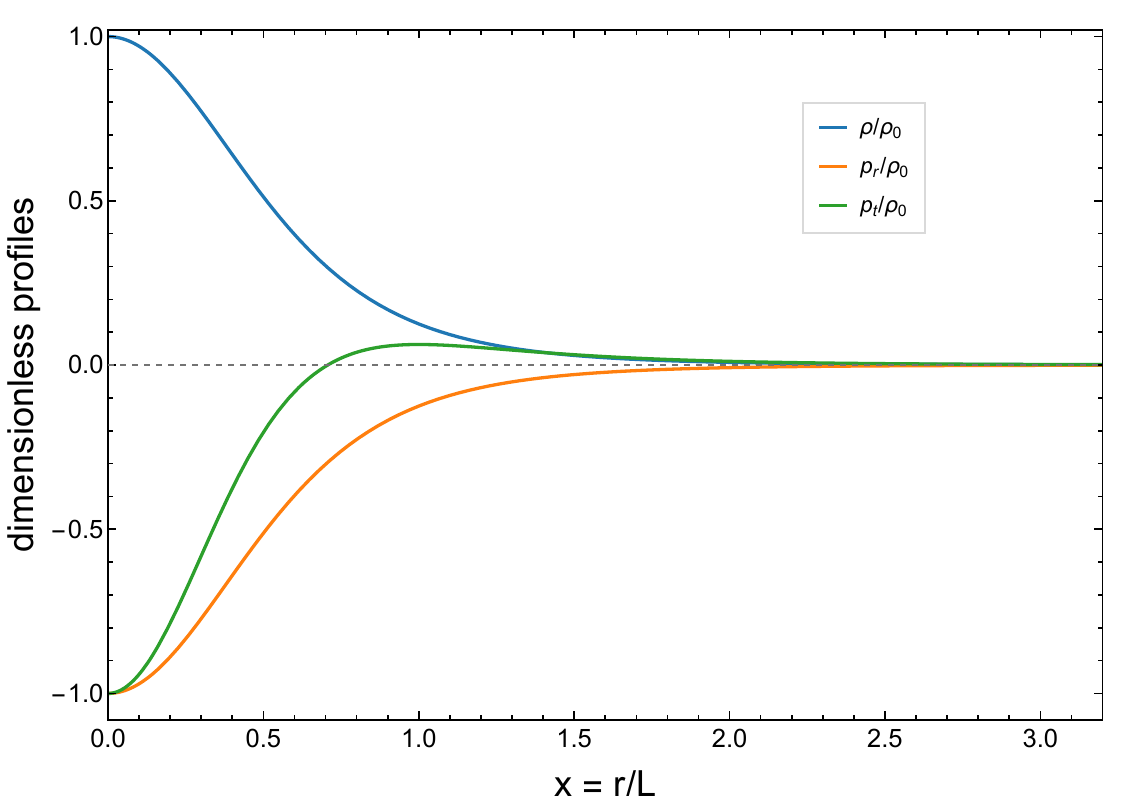}
\caption{Dimensionless density and pressures for the \(n=3\) solution. Near the centre the solution approaches an isotropic de Sitter core, \(p_r\simeq p_t\simeq -\rho_0\). By \(x\sim 3\), the asymptotic behaviour is already evident: \(\rho\sim r^{-6}\), \(p_r=-\rho\), and \(p_t\to 2\rho\).}
\label{fig:pressures}
\end{figure}

The energy conditions are correspondingly mild. The radial null energy condition is satisfied at the limiting value,
\begin{equation}
\rho+p_r=0,
\end{equation}
whereas the tangential null energy condition is satisfied,
\begin{equation}
\rho+p_t = Y K_Y \ge 0.
\end{equation}
Thus, the matter source does not violate the null energy condition in the angular directions, despite supporting a regular core. For the explicit $n=3$ branch, the weak energy condition is also satisfied, since $\rho\ge 0$, $\rho+p_r=0$, and $\rho+p_t\ge 0$. By contrast, the dominant energy condition is not satisfied globally, because $p_t/\rho\to 2$ as $r\to\infty$, and the strong energy condition is violated near the centre. This is not incidental, but rather the standard behaviour associated with regular black-hole interiors: a de Sitter-like core requires an effectively repulsive equation of state in the deep interior in order to prevent curvature blow-up. These properties are summarised in Fig.~\ref{fig:pressures}, which displays the dimensionless density and pressure profiles for the \(n=3\) solution. The figure highlights the isotropic de Sitter behaviour at the centre, the exact relation \(p_r=-\rho\), and the transition to an anisotropic but rapidly decaying outer region.

The scalar configuration carries topological rather than Gauss-law-like hair. Equation~\eqref{eq:fixedbranch} implies that the scalar fields define the standard hedgehog map from the sphere at spatial infinity into the internal two-sphere of fixed modulus~\cite{Barriola:1989hx}. The corresponding degree (winding number) is
\begin{equation}
\Qtop=\frac{1}{8\pi}\int_{S^2_\infty}\epsilon_{IJK}n^I\,\dd n^J\wedge \dd n^K=1.
\label{eq:Qtop}
\end{equation}
The solution therefore possesses genuine scalar hair, but this scalar hair is discrete and topological rather than continuous. In particular, there is no scalar Gauss-law charge analogous to electric charge. The continuous parameter $\rho_0$ of the exact family comes instead from the auxiliary three-form sector introduced in Eq.~\eqref{eq:action}; for the $n=3$ branch, Eq.~\eqref{eq:M3} shows that it is in one-to-one correspondence with the ADM mass. In this sense the family carries a continuous secondary hair parameter, but not a scalar charge independent of $M$; the non-trivial scalar structure itself is encoded in the global winding of the hedgehog configuration.

\subsection{Strong-field observables: photon sphere, shadow, ISCO, and eikonal ringdown}

Since the asymptotic correction to the metric is highly suppressed, weak-field observables remain essentially Schwarzschild-like when \(GM\gg L\). The clearest phenomenological differences therefore arise in the strong-field region, where circular null and timelike orbits govern the photon ring, the shadow, the inner edge of thin accretion disks, and the eikonal ringdown.

The unstable circular null orbit is determined by
\begin{equation}
rA'(r)-2A(r)=0.
\label{eq:phcond}
\end{equation}
Using the asymptotic expansion~\eqref{eq:Aasy}, one finds
\begin{equation}
\rph\simeq 3GM-\frac{32L^3}{27\pi G^2M^2}.
\label{eq:rph}
\end{equation}
Thus, at fixed ADM mass, the photon sphere lies slightly inside its Schwarzschild value. For a distant observer, the shadow radius is the critical impact parameter,
\begin{equation}
R_{\rm sh}=\frac{\rph}{\sqrt{A(\rph)}}
\simeq 3\sqrt{3}\,GM\left[1-\frac{16}{81\pi}\left(\frac{L}{GM}\right)^3\right].
\label{eq:Rsh}
\end{equation}
Hence the shadow is also slightly smaller than in Schwarzschild. The correction is tiny in the large-mass regime, but it is concentrated precisely in the strong-field region rather than in the post-Newtonian sector.

For timelike circular motion, it is convenient to denote the conserved specific angular momentum by \(\ell\). One then has
\begin{equation}
\ell^2(r)=\frac{r^3A'(r)}{2A(r)-rA'(r)},
\end{equation}
and the innermost stable circular orbit is determined by
\begin{equation}
\frac{d\ell^2}{dr}=0.
\end{equation}
In the large-mass expansion,
\begin{align}
\risco&\simeq 6GM-\frac{44L^3}{27\pi G^2M^2},
\label{eq:risco}\\
\OISCO&\simeq \frac{1}{6\sqrt{6}\,GM}\left[1+\frac{29}{81\pi}\left(\frac{L}{GM}\right)^3\right].
\label{eq:Oisco}
\end{align}
The ISCO therefore shifts inward, while the corresponding orbital frequency increases. In physical terms, the innermost stable part of the disk extends to slightly smaller radii than in Schwarzschild, so the characteristic orbital timescale is mildly blue-shifted.

Before turning to ringdown, it is worth stating the scope of the following estimate. We do not compute here the full coupled gravitational-triplet quasinormal spectrum. The formulae below should therefore be read as light-ring-based geometric-optics estimates tied to the background null orbit, rather than as a substitute for a dedicated perturbative calculation. In the standard test-field setting, eikonal quasinormal frequencies are often related to the orbital frequency and Lyapunov exponent of the unstable circular null geodesic~\cite{Cardoso:2008bp}. However, this correspondence is not universal: it can fail, or capture only part of the spectrum, for gravitational perturbations and for more general coupled perturbation sectors~\cite{Konoplya:2017wot,Konoplya:2022gjp}. With this caveat in mind, we use the background light ring as a diagnostic. The orbital frequency of the light ring and its Lyapunov exponent are
\begin{align}
\Oph&=\frac{\sqrt{A(\rph)}}{\rph}
\simeq \frac{1}{3\sqrt{3}\,GM}\left[1+\frac{16}{81\pi}\left(\frac{L}{GM}\right)^3\right],
\label{eq:Oph}\\
\lamph&\simeq \frac{1}{3\sqrt{3}\,GM}\left[1-\frac{32}{81\pi}\left(\frac{L}{GM}\right)^3\right].
\label{eq:lamph}
\end{align}
Accordingly,
\begin{equation}
\omega_{\ell n}^{\rm eik}\simeq \ell\,\Oph-i\left(n+\frac12\right)\lamph.
\label{eq:qnm}
\end{equation}
At the level of this light-ring-based eikonal estimate, the real part of the mode frequency is slightly larger than in Schwarzschild, whereas the damping rate is slightly smaller. The corresponding geometric-optics estimate thus points toward a marginally higher-frequency and longer-lived ringdown.

Eqs.~\eqref{eq:rph}--\eqref{eq:qnm} show that, for fixed ADM mass, the regular black hole is slightly more compact in the strong-field sense than Schwarzschild, with smaller characteristic orbit radii and correspondingly larger characteristic frequencies. At the same time, all deviations scale as $(L/GM)^3$, so they are strongly suppressed for macroscopic black holes and become relevant only in the photon-sphere and ISCO region, where the regular-core structure leaves its clearest imprint.

\section{Discussion and conclusions}
\label{sec:conclusions}

We have shown that General Relativity coupled to a constrained \(SO(3)\) scalar triplet and a non-propagating three-form sector admits an exact hedgehog branch with asymptotically flat geometrically regular black holes. The construction works as follows. A single scalar with explicit angular dependence cannot source an exactly spherical geometry, already at the level of the k-essence energy-momentum tensor. Once the angular dependence is absorbed into an internal index, however, the constrained triplet closes analytically: the modulus is frozen, the field equations dynamically enforce \(A=B\), and the Einstein sector reduces the metric to the first-order equation~\eqref{eq:massdef}. The role of the auxiliary three-form is to promote the overall density scale of the effective kinetic function to the integration constant \(\rho_0\), so that the exact regular branch forms a continuous family within one fixed theory.

Within the exact family generated by the functions~\eqref{eq:Kn}, the \(n=3\) member is the simplest asymptotically flat regular black-hole solution whose leading correction to Schwarzschild starts at order \(r^{-4}\). In this sense, it is less monopole-like than the standard global-monopole geometry and less Reissner--Nordstr\"om-like than the \(n=2\) member of the same family. The regular core is de Sitter, while the far field remains unusually close to Schwarzschild.

The scalar hair carried by the solution is genuine, but it is topological rather than Coulomb-like. The relevant invariant is the winding of the hedgehog map \(S^2_\infty\to S^2_{\rm int}\), not a new continuous scalar charge. The continuous parameter \(\rho_0\) of the exact family comes from the auxiliary three-form sector, not from a Gauss law of the triplet fields. For the \(n=3\) branch one has \(M=\pi^2\rho_0L^3/4\), so \(\rho_0\) is secondary to the ADM mass: it parameterises the family, but it is not an independent scalar charge at infinity. In this sense the solutions carry continuous secondary hair, but not a scalar charge.

The roles of the Lagrange multiplier and of the three-form should both be interpreted in this effective sense. The former freezes the heavy radial mode and leaves the Goldstone directions, while the latter is a minimal covariant device that turns the overall density scale into an integration constant without introducing new local propagating degrees of freedom. In this sense, the auxiliary sector should be viewed as a device for generating the continuous family parameter of the exact solutions, not as an additional source of local dynamical hair. The construction is therefore most naturally read as an effective theory for the angular Goldstone directions rather than as a fundamental ultraviolet completion.

Regularity itself requires a qualification. The black holes constructed here are geometrically regular: the mass function, metric coefficients, and polynomial curvature invariants remain finite at the centre, and the core is de Sitter. At the same time, the matter sector is less regular than the geometry. On the exact branch one has \(Y=\eta^2/r^2\), so the kinetic invariant diverges at the central point and the order parameter \(\Phi^I=\eta n^I\) is not smooth there because \(n^I=x^I/r\) is undefined at \(r=0\). The regularity mechanism is therefore one in which \(K(Y)\) and the stress tensor remain finite even though the underlying angular order parameter is not smooth at the origin. The precise claim of the paper is thus geometric regularity rather than complete smoothness of the matter fields.

If one insists on full matter smoothness at the centre, the fixed-modulus truncation is too restrictive. A dynamical-modulus solution with \(H(r)=h_1 r+O(r^3)\) would instead give a finite central value \(Y=\frac32 h_1^2+O(r^2)\), so the matter sector could remain regular together with the geometry. Such solutions lie outside the exact branch studied here and would require solving the full coupled system with a genuine radial mode.

From the black-hole point of view, the $n=3$ branch has a simple thermodynamic structure at the level of the exact static backgrounds studied here. The horizon structure contains horizonless, extremal, and two-horizon configurations, with the extremal point marking the onset of the black-hole phase. The temperature vanishes at extremality, rises to a single maximum, and then falls back toward the Schwarzschild regime. Equivalently, the heat capacity is positive close to extremality, diverges at the temperature maximum, and becomes negative for sufficiently large masses. The branch is therefore locally thermodynamically stable near extremality but thermodynamically unstable in the large-mass regime.

The clearest effects are therefore concentrated in the strong-field region: the photon sphere and the ISCO move inward, the corresponding orbital frequencies increase, and the light-ring estimate for the real part of the eikonal frequency is shifted upward while the associated damping rate is shifted downward. In this sense the model shifts its most visible signatures toward the photon sphere and accretion region rather than toward post-Newtonian scales. A full perturbative analysis is therefore required to establish the actual quasinormal spectrum of the coupled system, especially because the eikonal light-ring correspondence is not universal for gravitational or coupled perturbation sectors~\cite{Konoplya:2017wot,Konoplya:2022gjp}.

A first priority is a full analysis of linear perturbations and the associated question of dynamical stability. For regular black holes this issue is essential, since regularity of the background does not by itself guarantee a healthy perturbation spectrum. In the present model the problem is also qualitatively different from the nonlinear-electrodynamic case: the background is supported by a constrained scalar triplet with genuine angular structure, so even the axial sector need not reduce to the vacuum Regge--Wheeler problem and may already involve non-trivial couplings between metric and matter perturbations. More generally, odd-parity analyses of hairy black holes and scalar-tensor solutions have repeatedly proved to be a sharp probe of stability, strong coupling, and the effective propagation structure of perturbations~\cite{DeFelice:2024seu,Watabe:2003ak,Ogawa2016,Minamitsuji2022,Langlois2022Axial,Charmousis2025Axial}. Another natural extension is to relax the fixed-modulus constraint and restore a dynamical radial mode. Such a completion could preserve the geometrical regularity of the solutions while improving the smoothness properties of the matter sector at the centre. The exact solutions obtained here provide a simple analytic starting point for these questions.

\section*{Acknowledgements}
The work of S.B. is supported by the Institute for Basic Science (IBS-R018-D3). The author would like to thank Theodoros Nakas, Pavel Petrov, Eduardo Guendelman, Emil Nissimov, and Svetlana Pacheva for useful discussions.

\bibliographystyle{utphys}
\bibliography{references}

\end{document}